# Extensible Validation Framework for DSLs using MontiCore on the Example of Coding Guidelines

Christian Berger, Bernhard Rumpe, and Steven Völkel

Rheinisch Westfälische Technische Hochschule (RWTH) Aachen
Department of Computer Science 3
Ahornstraße 55, 52074 Aachen, `www.se-rwth.de`

**Abstract.** Unit test environments are today's state of the art for many programming languages to keep the software's quality above a certain level. However, the software's syntactic quality necessary for the developers themselves is not covered by the aforementioned frameworks. This paper presents a tool realized using the DSL framework MontiCore for automatically validating easily extensible coding guidelines for any domain specific language or even general purpose languages like C++ and its application in an automotive R&D project where a German OEM and several suppliers were involved. Moreover, it was exemplary applied on UML/P-based sequence charts as well.

## 1 Introduction

The most popular and sophisticated methods for assuring the quality of software today are unit test environments which are, especially for the Java programming language, well combined with popular integrated development environments (IDE) like Eclipse. For developers, especially those who are new to a project, the source code's syntactic quality is also important to reduce the time necessary to understand the code and the concepts behind it. Coding guidelines (CGLs) like [1–5] are available to be used in projects, but they are often regarded as recommendations rather than as regulations and their fulfillment is not checked regularly due to missing appropriate tools. However, coding guidelines improve the source code's quality implicitly by warning the developer in case of complex statements or poor names for variables.

Tools for checking the source code's conformance to a set of coding guidelines are for languages other than Java hardly available. Especially for C++, besides C and Matlab/Simulink the major language for implementing software for embedded systems, tools for automatically checking the source code are only a few unavailable. The main reason is the complexity of the C++ language itself. Using the preprocessor concept, nearly the entire language could be redefined or extended.

This paper presents a tool for checking any domain specific language's conformance to a set of easily extendable guidelines using a MontiCore generated parser and symbol table and is structured as follows. Sec. 2 lists some related work with respect to tools for validating C++ guidelines and discusses differences to our approach. In Sec. 3 the DSL framework MontiCore and its concepts for developing domain specific language processing tools is presented. The following Sec. 4 outlines the generic validation framework realized by using MontiCore. Its application to C++ in an automotive R&D project and results of using this tool are presented in Sec. 5. In Sec. 6 its application to the domain specific language for sequence charts from the UML/P is demonstrated. A conclusion is given in the last section.

## 2 Related Work

Tools for checking the source code's conformance to coding guidelines for C++ are only a few available. Most commercial tools like the one from Parasoft[1] use the compiler front-end provided by Edison Design Group[2].

Another commercial tool is QA-C++ offered by QASystems. This tool also performs static code analysis comparable to Parasoft's one and to VF.

Cxxchecker[3] is an open source tool for validating coding guidelines for C++. This tool only checks naming conventions, no public data members, the existence of a copy constructor if pointer attributes exist and no public constructors in abstract classes.

As aforementioned, some compilers like GNU G++ and Microsoft Visual C++ offer the possibility to enable specific warning levels to avoid potentially malicious code. Furthermore, regular expressions can be used to perform rudimentary checks for naming conventions.

There are three main differences of these approaches compared to our work described in this paper. First, we provide a language-independent framework which can serve as a basis to develop a validation tool. According to our experience, the effort needed to implement such a tool decreases significantly. Second, we developed an *extensible* framework to check coding guidelines for C++. Users can easily define their own rules which can then be hooked in into the framework. And third, we have an integrated tool which includes means to define concrete syntax, abstract syntax, symbol tables, code generators etc. This integration permits a seamless development whereas using different tools for the development of these artifacts often leads to integration problems.

## 3 MontiCore – A DSL-Framework

MontiCore (e.g., [6–9]) is a framework for agile development of textual domain-specific and general purpose languages developed at the Software Engineering Group, RWTH Aachen. It combines classical grammar-based concepts with today's meta-modeling approaches and permits an integrated definition of abstract and concrete syntax. Due to its comprehensive functionalities and especially its extensibility, we have chosen to use MontiCore as basis for our generic validation framework. However, as MontiCore is subject of several publications in literature, we will only briefly describe its characteristics in this paper.

---

[1] http://www.parasoft.com
[2] http://www.edg.com
[3] https://gna.org/projects/cxxchecker

### 3.1 Integrated Definition of Abstract and Concrete Syntax

MontiCore uses one single format similar to EBNF to define both concrete and abstract syntax of a language. Fig. 1 shows an excerpt of our C++ grammar.

```
──────────────────────── MontiCore-Grammar ────────────────────────
1 token IDENT = ('a' ... 'z' | 'A' ... 'Z')+;
2
3 Type = Class | Enum;
4
5 Class = "class" name:IDENT "{" ClassBody "}";
6
7 Enum = "enum" name:IDENT "{" EnumBody "}";
8
9 ...
```

**Fig. 1.** Excerpt of our C++ grammar.

The first line introduces a terminal `IDENT` which defines identifiers to be used as names of classes, variables, or methods[4]. Line 3 defines a nonterminal `Type` which can be either derived to a class or an enum. Furthermore, lines 5 and 7 define the structure of classes and enums, respectively.

From this grammar, MontiCore generates both a parser and a lexer which can be used to process C++ files. Furthermore, abstract syntax classes are generated which will be instantiated as an abstract syntax graph (ASG) by the parser. This ASG serves as a starting point for further processors such as our generic validation framework.

### 3.2 Framework-based Approach for Language Processors

MontiCore offers a comprehensive framework which can be used as a basis for the development of language processors. Beyond the aforementiond parsers, lexers, and abstract syntax classes which are always language dependent, the framework itself concentrates on cross-cutting, language-independent functionalities. Typical examples are error handling, file creation, or structuring of the processing workflow. Especially the latter property is of special interest as it permits extensibility of our generic validation framework.

---

[4] Note that the grammar is an excerpt only. Therefore, IDENTs have been simplified.

As stated above, the central entity for every language processor is the ASG. The ASG and other useful information like a symbol table, the filename, or the content of the file are stored in a class called `DSLRoot`. Furthermore, MontiCore provides the concept of workflows which are used to encapsulate algorithms on `DSLRoots` such as parsing, symbol table establishment, pretty printing, or - like in our case - checking coding guidelines. The results of these algorithms are again stored in the root and are thus available for other workflows.

Another concept supported by MontiCore are visitors. They are mainly used in order to traverse the ASG based on the structure of the language's grammar. Typical applications are code generators, symbol table builders, or - as we will show later - coding guideline checkers.

### 3.3 Further Concepts of MontiCore

Beyond the concepts we described in the former subsections, MontiCore offers means for all parts of the development of language specific tools. For the sake of space, we will only enumerate the main functionalities. For more information we refer to [10].

- *Support of meta-modeling concepts.* MontiCore supports common meta-modeling concepts in its abstract syntax definition (e.g., inheritance, interfaces, and bidirectional associations).
- *Compositionality of languages.* MontiCore offers two concepts for compositionality on language level. Inheritance can be used to inherit from existing languages and to specify the delta only. Embedding supports combination of languages (e.g., Java with embedded SQL) at runtime.
- *Attribute grammar system.* The MontiCore framework is equipped with an attribute grammar system which can be used for different purposes like code generation or type analysis. This attribute grammar system respects compositionality of languages.
- *Editor generation.* MontiCore is able to generate language-specific editors as Eclipse-plugins with different comfort functionalities (e.g., syntax highlighting, outlines, and autocompletion). The editor generation mechanisms respect compositionality of languages.

## 4 A Generic and Extensible Validation Framework

Our approach for a generic and extensible validation framework (VF) bases to some extent on [11]. As a start, overall design criteria and architectural considerations are discussed for using this framework for different domain specific languages or general purpose languages like C++.

### 4.1 Architectural Drivers and Design Considerations

The most important design considerations were extensibility and reusability for different languages. Therefore, the VF must not operate directly on the syntax of a concrete language but should rather use an intermediate representation. We decided to use the abstract syntax graph as provided by our MontiCore framework for this representation.

Since we decided to use the ASG itself as the representation for any language the visitor design pattern could be easily applied to it for traversing all nodes and querying necessary information like attributes or associations. For avoiding every validation rule (VR) like a coding guideline to be implemented as a visitor and running through the ASG several times which is very expensive in terms of runtime complexity, we decided to use one generic visitor for traversing all nodes only once and every VR registers itself at the visitor to get notified about `ASGNode`s. Furthermore, it should be possible to select the VR to be used during the visitor's traversal as well as to configure the VR itself easily without touching the implementation. Thus, a simple configuration file defines which VRs should be applied to the ASG and sets up some parameters.

Another aspect was the possibility to integrate the VF into a continuous integration system for automatically validating the source code triggered by modifications. Additionally, this requirement demands a reporting interface for collecting the output of all matching rules applied to the source code in an independent representation to be used in different reporting contexts like a browseable page in a Web portal or for email notification containing only a summary of the results.

### 4.2 General Architecture

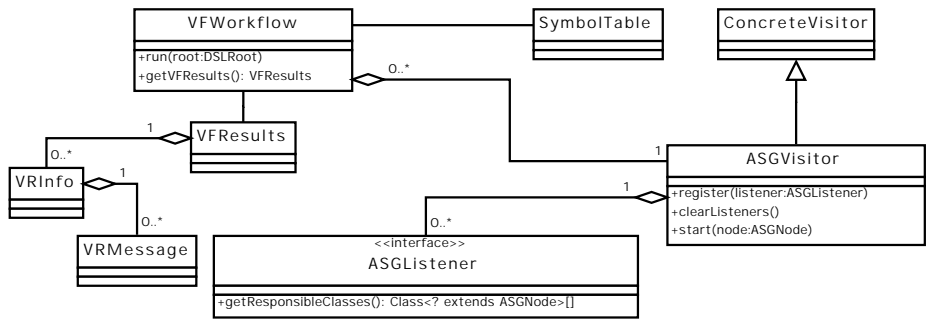

**Fig. 2.** Architecture for the Validation Framework.

As shown in Fig. 2, the topmost two classes `VFWorkflow` and `ConcreteVisitor` are provided by the MontiCore framework as described above. They realize the traversal through an ASG constructed from an instance of a given grammar. The abstract class `ASGVisitor` is the VF's main visitor class and must be derived once for every language to handle specific nodes from the ASG. At this class, subclasses of `ASGListener` implementing a specific validation rule like a coding guideline are registered for getting notified if matching nodes in the ASG are visited. If necessary, a concrete VR can use the symbol table also provided by MontiCore to query a variable's visibility for example.

All validation rules report their results to `VFResults` by using a `VRInfo` instance per applied coding guideline. Information about successful matches of a specific validation rule are collected using `VRMessages` containing information about the file, row and column where the match succeeded as shown in the VF's workflow depicted in Fig. 3. The root object `VFResults` is serialized to an XML file using XStream[5] which can be processed further using an XSL transformation as shown in Sec. 5.4.

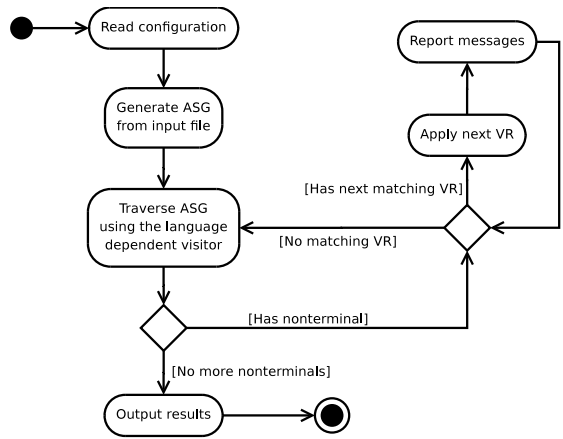

**Fig. 3.** Validation Framework's workflow for checking an ASG.

## 5 Application in an Automotive R&D-Project

For testing the developed framework with real data, we applied it within an automotive R&D-project carried by a German OEM. Furthermore, selective use of the well-known Motor Industry Software Reliability Association (MISRA) rules in its elaborated version [12] for coding guide lines was required. Thus, we had to implement the chosen rules to apply them automatically on the source code. This section describes shortly the project itself and describes the C++ grammar we implemented using our MontiCore framework. Afterwards, an implementation of a selected validation rule is presented as well as a report produced by the validation framework.

---
[5] `http://xstream.codehaus.org`

### 5.1 Project Settings

The goal of the project was the development of an automatically driving vehicle for highways [13]. The system architecture consisted on the one hand of an environment perception and driving decision component using different sensors to detect the car's surroundings like other vehicles or obstacles beside the road. This component was realized using C++. On the other hand, parameters for all actuators were computed by a controller component realized using Matlab/Simulink. Both components communicated using the CAN bus.

The former component was tested interactively by the developers as well as automatically by a continuous integration system on every single commit to the versioning system Subversion [14]. Therefore, the coding guidelines validation tool should be part of this automatic workflow as well.

### 5.2 The C++ Grammar

The regular compiling process for C++ source code starts with the preprocessing stage[6]. In this stage all preprocessor directives are resolved and `#include` statements are embedded directly into a source file. For the sake of simplicity, our implementation starts right after this stage.

```
────────────────────────── MontiCore-Grammar ──────────────────────────
1 ...
2 Simple_type_specifier =
3   {symbolHelper.qualifiedItemIsOneOf(this,
4     EnumSet.<QualifiedItem>of(QualifiedItem.qiCtor, QualifiedItem.qiType))
5   }?
6   QualifiedType:Qualified_type
7 | ... // more rules
```

**Fig. 4.** Example for a semantic predicate.

The C++ language itself is a context-dependent grammar. An overview of ambiguities can be found in [15]. Thus, it cannot be parsed using an $LL(k)$ or $LR(k)$ parser without additional information from the context by either syntactical predicates using further tokens or semantic predicates using information from the symbol table. As shown in Fig. 4, the production rule `Simple_type_specifier` queries the symbol table to check if the following type is already known.

The complete MontiCore-grammar for the C++ contains 155 production rules in 2,652 lines.

---
[6] The GNU C++ compiler produces this output using the command `g++ -E mySourceFile.cpp`.

### 5.3 Use of MISRA for C++ in the Validation Framework

As mentioned before, MISRA coding guidelines were adopted by the aviation industries and are mandatory for safety critical systems. MISRA rules are divided into *SHOULD*, *SHALL* and *WILL*. The first two classes are obligatory, whereas the second class allows exceptions. The third one can be applied but the source code's conformance to it is not verified.

| | |
|---|---|
| `ConstructorChecker` | Validation of a specific order of constructors, destructor and other methods |
| `DestructorChecker` | Asserting a virtual destructor |
| `EnumChecker` | Validation ISO-like enum-declarations |
| `ExpressionChecker` | Checking for left-factorized expressions |
| `ExpressionAssignmentChecker` | Validation for assignments in selection statements |
| `FlowControlChecker` | Checking for flow control (`goto`s, `break`s, `label`s) |
| `FunctionChecker` | Various validations for function and method usage (e.g., naming, maximum number of lines...) |
| `IdentifierChecker` | Ensuring for unique identifiers |
| `IfChecker` | Checking for braces in if-statements |
| `InitializedVariableChecker` | Checking the initialization of local variables |
| `InterfaceChecker` | Ensuring that all public methods from a class are declared in interfaces used by this class |
| `MemoryChecker` | Validation for correctly freed memory |
| `NamingConventionChecker` | Checking for correct naming convention (e.g., avoiding Hungarian prefixes...) |
| `NamespaceChecker` | Ensuring correct namespace usage |
| `SingleLetterVariableChecker` | Searching for variable names containing only one letter |
| `SwitchChecker` | Checking for braces in switch-case-statements |
| `SymbolOrderChecker` | Validation for compact symbol declaration (e.g., at the beginning) |
| `TypeDefChecker` | Validation for correct naming of `typedef`s |

**Table 1.** List of implemented coding guidelines.

For implementing MISRA guidelines using the VF, `ASGVisitor` must be derived as mentioned before. Thus, we implemented `ASGCPPVisitor` which delegates visited nodes in the ASG to all registered listeners for a specific node. Altogether, 18 validation rules as subclasses of `ASGListener` were implemented as described in the Table 1.[7]

---

[7] Compilers might also perform some of these checks using appropriate warning levels.

In the following, the implementation for `InterfaceChecker` is presented. For validating the coding guideline that all declared public methods must be provided by at least one interface, the source code shown partly in Fig. 5 was used.

```java
public void notifyVisit(ASGClass_decl_or_def d) {
  Scope s = SymbolCache.getInstance().getScope(d);
  ClassScope c = ((ClassScope) s);
  ClassBinding cB = c.getBinding();

  List<FunctionBinding> implFs = new LinkedList<FunctionBinding>();

  for(ClassBinding iC : cB.getInheritedClasses()) {
    if (iC.hasOnlyInterfaceMethods()
        && cB.getSpecifierOfInheritedClass(iC)
            .contains(Specifier.PUBLIC)) {
      for(FunctionBinding fB : iC.getAllFunctions()) {
        if (fB.hasSpecifier(Specifier.PUBLIC))
          implFs.add(fB);
      }
    }
  }
  for(FunctionBinding fB : cB.getAllFunctions()) {
    boolean notInInterface = true;
    if (fB.hasSpecifier(Specifier.PUBLIC)) {
      for(FunctionBinding b : implFs) {
        if (fB.equalSignature(b))
          notInInterface = false;
      }
    } else {
      notInInterface = false;
    }
    if (notInInterface)
      addMessage(d, "Class " + cB.getName() + " has public functions"
        + " not declared in interfaces: " + fB.printSignature());
  }
}
```

**Fig. 5.** Example for validating correct interface usage.

In line 2-4, information about the current node is retrieved from the symbol table. The following lines 8-17 collect all methods from the current class that are inherited from interfaces. Finally, the last lines 18-31 verify that all public methods are declared in super classes implemented by the current class.

### 5.4 Results

The VF was integrated into the continuous integration system (CIS) of the automotive R&D project. Thus, the source code's conformance to the previously defined coding guidelines could be verified automatically. For providing an easily accessible report about the source code's quality, the results were embedded into the developer's Web portal Trac[8] as shown in Fig. 6 using an XSL transformation.

**Fig. 6.** Automatically generated report in a developer's Web portal.

During the project, for about 100,000 lines of code were analyzed and totally 5,338 VF warnings were found. Nearly 8.6% were classified as highly critical VF warnings like assignments in boolean expressions or misuse of memory handling. For about 29.5% are classified as warnings of medium criticality like uninitialized variables and their use in logical expressions. The remaining 61.9% were non critical warnings like layout or design violations (e.g., not all public methods are inherited from interfaces) which could be fixed automatically to be conform to predefined layout rules. All critical warnings have been reviewed manually by the developers to fix potentially malicious code. Furthermore, due to the tool's integration into the CIS, a regular report about the source code's quality could be provided automatically to check the source code's quality over time.

---

[8] http://trac.edgewall.com.

## 6 Application on UML/P

To test the eligibility for other languages, we decided to implement a simple coding guidelines validation tool for sequence diagrams based on UML/P[16]. The criteria listed in Table 2 were tested[17].

| | |
|---|---|
| `TriggerChecker` | Checking for stereotype ≪`trigger`≫ at the beginning of an interaction |
| `NoCallToTestDriverChecker` | Validation that the object under test does not call the test driver itself |

**Table 2.** List of implemented coding guidelines for UML/P sequence charts.

In Fig. 7, a sequence chart used as test case specification is shown. In line 9 the mandatory ≪`trigger`≫ statement is omitted. In line 21, the system under test tries to call the test runner which is not permitted at all.

────────────────────── UML/P ──────────────────────

```
 1 sequencediagram librarytest {
 2   object test:LibraryTest;
 3   object library:Library;
 4   object librarian:Librarian;
 5   object client:Client;
 6   object request:Request;
 7   object book:Book;
 8   {
 9     test -> librarian : setup(); // <<trigger>>  missing.
10     test -> librarian : <<trigger>> startService();
11
12     librarian -> client : startBorrowing();
13     {
14       client -> librarian : requestBook(request);
15       librarian -> library : requestBook(request);
16       librarian <- library : return book;
17       client <- librarian : return book;
18     }
19     librarian <- client : endBorrowing();
20
21     librarian -> test : finish(); // No calls to test runner allowed.
22   }
23 }
```

────────────────────────────────────────────────────

**Fig. 7.** Malicious sequence chart.

For using the VF to validate instances of this DSL, the `ASGVisitor` had to be derived as well as the `ASGListener` to implement the rules. Everything else like the reporting interface could be simply reused. Since the UML/P itself is part of the MontiCore framework, it was pretty easy to apply the VF on the ASG.

Besides C++ and sequencs diagrams, we developed several other validation tools for other languages. Thiese include other UML-languages (class diagrams, state charts, object diagrams), Java, the MontiCore grammar format itself, and OCL. All these validation tools are based on the framework we described in this paper. This shows the applicability of the language independent validation framework.

# 7 Conclusion

In this work we have presented a framework for domain specific languages as well as general purpose languages for defining validation rules like coding guidelines. Besides design concepts and its generic architecture we have shown its application in a real project from the automotive domain. By applying our tool we discovered several potentially malicious sections in the R&D project's source code and improved its quality. Since the tool was combined with a continuous integration system the validation of source code was conducted automatically and the reports were delivered to the developers using the popular web portal Trac. Furthermore, we have shown the generic nature of the VF by successfully applying it to sequence charts from the UML/P and other languages.

# References

1. Meyers, S.: Effective C++: 55 Specific Ways to Improve Your Programs and Designs. Addison Wesley Professional (2005)
2. Henricson, M., Nyquist, E.: Industrial Strength C++. Prentice Hall (1997)
3. Sutter, H., Alexandrescu, A.: C++ Coding Standards. Addison Wesley (2005)
4. GNU Website. http://www.gnu.org/prep/standards/standards.html
5. Google C++ Style Guide Website http://google-styleguide.googlecode.com
6. Krahn, H., Rumpe, B., Völkel, S.: Integrated Definition of Abstract and Concrete Syntax for Textual Languages. In: Proceedings of Models 2007. (2007)
7. Krahn, H., Rumpe, B., Völkel, S.: Efficient Editor Generation for Compositional DSLs in Eclipse. In: Proceedings of the 7th OOPSLA Workshop on Domain-Specific Modeling 2007. (2007)
8. Krahn, H., Rumpe, B., Völkel, S.: Monticore: Modular development of textual domain specific languages. In: Proceedings of Tools Europe. (2008)
9. Grönniger, H., Krahn, H., Rumpe, B., Schindler, M., Völkel, S.: Monticore: a framework for the development of textual domain specific languages. In: 30th International Conference on Software Engineering (ICSE 2008), Leipzig, Germany, May 10-18, 2008, Companion Volume. (2008) 925–926
10. MontiCore Website http://www.monticore.de

11. Witt, S.: Entwurf und Implementierung einer erweiterbaren Qualitätssicherungsschnittstelle für Codierungsrichtlinien im automobilen Forschungsumfeld für das Framework MontiCore am Beispiel der Sprache C++. Master's thesis, Technische Universität Braunschweig (2007)
12. AT&T Research website `http://www.research.att.com/~bs/JSF-AV-rules.pdf`: Joint Strike Fighter Air Vehicle (December 2005)
13. Weiser, A., Bartels, A., Steinmeyer, S., Schultze, K., Musial, M., Weiß, K.: Intelligent Car – Teilautomatisches Fahren auf der Autobahn. In Gesamtzentrum für Verkehr Braunschweig e.V., ed.: AAET 2009 – Automatisierungssysteme, Assistenzsysteme und eingebettete Systeme für Transportmittel. Volume 10. (February 2009) 11–26
14. Bartels, A., Berger, C., Krahn, H., Rumpe, B.: Qualitätsgesicherte Fahrentscheidungsunterstützung für automatisches Fahren auf Schnellstraßen und Autobahnen. In Gesamtzentrum für Verkehr Braunschweig e.V., ed.: AAET 2009 – Automatisierungssysteme, Assistenzsysteme und eingebettete Systeme für Transportmittel. Volume 10. (February 2009) 341–353
15. Willink, E.D.: Meta-Compilation for C++. PhD thesis, University of Surrey (2001)
16. Rumpe, B.: Agile Modellierung mit UML : Codegenerierung, Testfälle, Refactoring. Springer, Berlin (August 2004)
17. Fraikin, F.: Entwicklungsbegleitendes Testen mittels UML-Sequenzdiagrammen. PhD thesis, Technische Universität Darmstadt (2003)